\documentclass[asr]{copernicus2}  
\usepackage{fixltx2e} 
\usepackage{journalabbrevs}

\definecolor{dgreen}{rgb}{0.0,0.5,0.0}
\definecolor{dpurple}{rgb}{0.75,0.0,0.75}
\usepackage[colorlinks,citecolor=dgreen,linkcolor=dpurple,urlcolor=blue]{hyperref}

\usepackage{doi}

\graphicspath{{./figs/}}  

\newcommand{\windunit}{\,\unit{m\,s^{-1}}}
\newcommand{\trendunit}{\,\unit{m\,s^{-1}\,decade^{-1}}}

\title{European wind variability over 140 yr}
\author{P. E. Bett}
\author{H. E. Thornton}
\author{R. T. Clark}
\affil{Met Office Hadley Centre, Exeter, UK}
\runningauthor{Bett, Thornton, Clark}
\runningtitle{European wind over 140 yr}
\correspondence{Philip Bett\\ (philip.bett@metoffice.gov.uk)}

\begin{document}

\abstract{ We present initial results of a study on the variability of
  wind speeds across Europe over the past 140 yr, making use of the
  recent \emph{Twentieth Century Reanalysis} data set, which includes
  uncertainty estimates from an ensemble method of reanalysis.  Maps
  of the means and standard deviations of daily wind speeds, and the
  Weibull-distribution parameters, show the expected features, such as
  the strong, highly-variable wind in the north-east Atlantic.  We do
  not find any clear, strong long-term trends in wind speeds across
  Europe, and the variability between decades is large.  We examine
  how different years and decades are related in the long-term
  context, by looking at the ranking of annual mean wind speeds.
  Picking a region covering eastern England as an example, our
  analyses show that the wind speeds there over the past $\sim 20$
  yr are within the range expected from natural variability, but do
  not span the full range of variability of the 140-yr data set.
  The calendar-year 2010 is however found to have the lowest mean wind
  speed on record for this region.  }

\maketitle


\introduction

Knowing the form of the wind speed distribution is of critical
importance when assessing the wind energy potential at a site.
Typically, when wind farm developers or investors consider a site,
they assess it using (at best) the past 20--30 yr, with data from
direct observations, NWP models, and reanalyses.  These recent decades
reflect our personal experience of wind speeds, but they do not show
the longer-term historical context. Understanding whether the most
recent decades were more or less windy than normal, or if there are
any significant long-term trends, is key to understanding the range of
possible future windspeeds we might experience over the coming $\sim
5$ yr, or over the lifetime of a wind farm ($\sim 25$ yr).
This information is important not just for managing wind farms, but
also for planning investment in future wind energy projects.

In this study, we show wind speed distributions for Europe over 140
yr (1871--2010), utilising the \emph{Twentieth Century Reanalysis}
data set \citep[20CR,][]{Compo2011}.  This reanalysis incorporates
observations of sea-level pressure and surface pressure alone, with
sea-surface temperature and sea-ice concentration data used as
boundary conditions.  The reanalysis used an experimental coupled
atmosphere--land version of the NOAA NCEP Global Forecast Model, ran
with 56 ensemble members.  This allows an estimate of the
uncertainties present due to episodes with less data (for example due
to the reduction in Atlantic shipping during the World Wars, or the
overall reduction in observations as one looks further back in time).
We use the daily-mean wind speeds at the near-surface pressure level
where $P/P_\mathrm{surface} = 0.995$, on a regular lat--lon grid of
resolution $2\degree \times 2\degree$.  Other studies of European wind
speed climatologies have used data sets that assimilate more
observations or are at higher resolution.  For example, the studies of
\cite{Kiss2008}, \cite{Pryor2006}, and \cite{Siegismund2001} use the
ERA-40 \citep{Uppala2005} and NCEP/NCAR reanalyses
\citep{Kalnay1996NCEPNCAR}, which both span $\sim 50$ yr.  The key
feature of the use of the 20CR in our study is the unprecedented
length of the time series, which allows us to compare decadal-scale
fluctuations in the speed and variability of the wind.

In this short article, we present an initial sample of our results. We
reserve a more full analysis for a later publication.


\section{Results}

\subsection{Wind speed and variability}
The simplest way of describing the long-term wind speed distribution
over Europe is to map the long-term mean and standard deviation of the
daily wind speeds.  The long-term average is simply the mean of the
daily wind speeds from all days from all ensemble members.  For the
standard deviation, we need to take account of the structure of the
20CR ensemble: the 56 members of the reanalysis were run as a series
of independent ``streams'', each covering consecutive 5-yr
periods\footnote{Streams 16 and 17 actually have lengths of 6 and 4
  yr respectively, and stream 27 consists of 10 yr (1999--2010).
  We treat the whole series as 28 equal periods of 5-yr, for
  simplicity.}  \citep[see Table III in][]{Compo2011}. The result of
this is that a time series from an ensemble member is only temporally
continuous for 5 yr, and aggregating over any period longer than
this risks mixing interannual variability with ensemble
member-to-member variability.  Therefore, for the long-term standard
deviation, we first calculate standard deviations in 5-yr periods
for each ensemble member.  We then take ensemble means, giving us a
single standard-deviation time series in 5-yr steps.  These are then
aggregated into a single long-term ensemble-mean value,
using\footnote{Note that equation (\ref{eq:lgpopstdev}) is for the
  population standard deviation, which we use as we are aggregating
  data covering the complete time period, rather than estimating
  $\sigma$ from a sample of years.}
\begin{equation}
  \label{eq:lgpopstdev}
\sigma^2 = \frac{\sum_i N_i \left(\sigma^2_i+\bar{U}_i^2\right)}
                {\sum_i N_i} 
                - \bar{U}^2,
\end{equation}
where $N_i$ is the number of days in the 5-yr period $i$, the
ensemble-means of the 5-yr mean and standard deviations of the daily
wind speeds are $\bar{U}_i$ and $\sigma_i$ respectively, and the
long-term mean wind speed is $\bar{U} \equiv \left(\sum_i \bar{U}_i
\right) / 28$.

\begin{figure} 
  \vspace*{2mm}
  \begin{center} 
    \includegraphics[width=0.8\columnwidth]{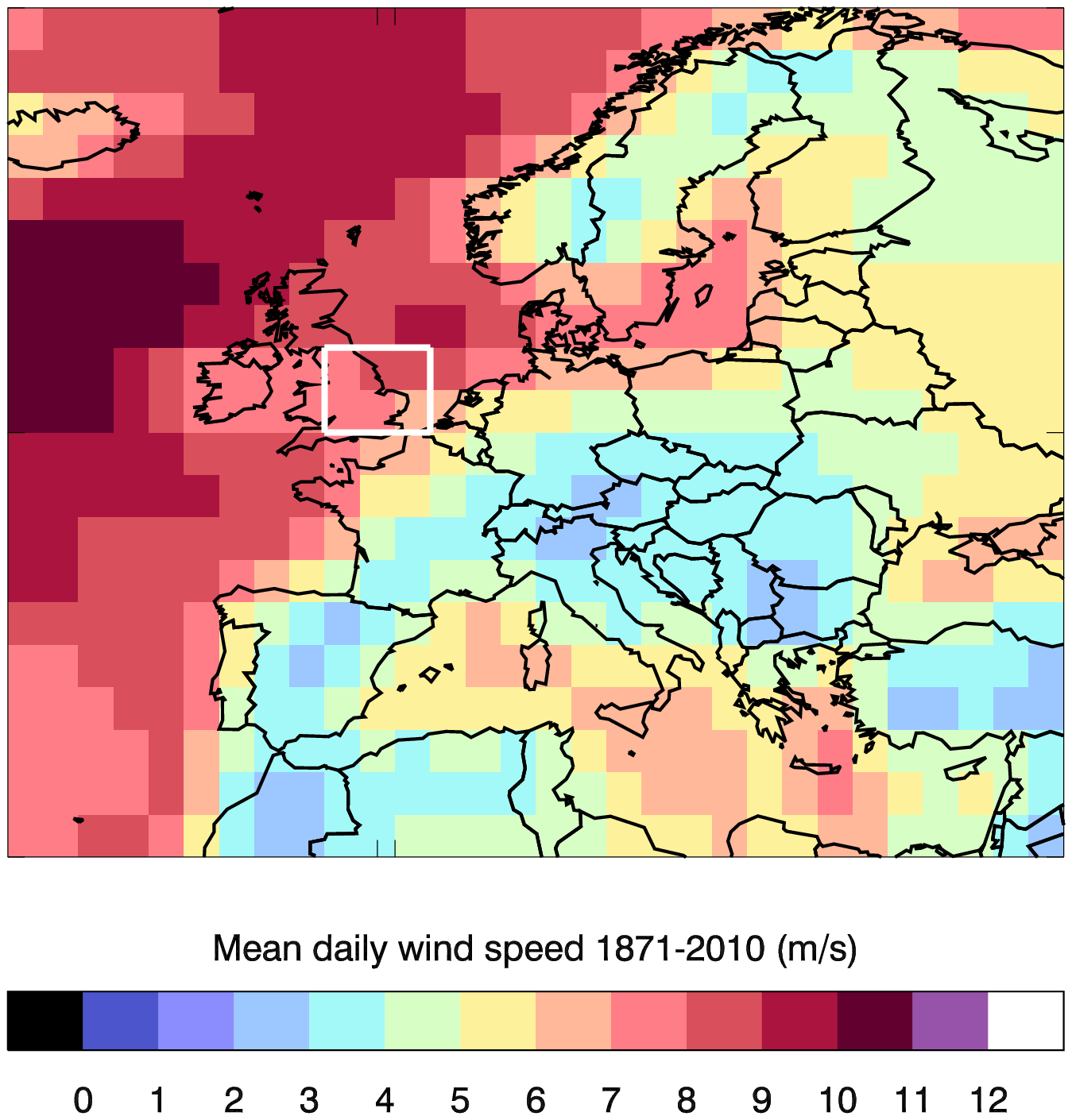} \\
    \includegraphics[width=0.8\columnwidth]{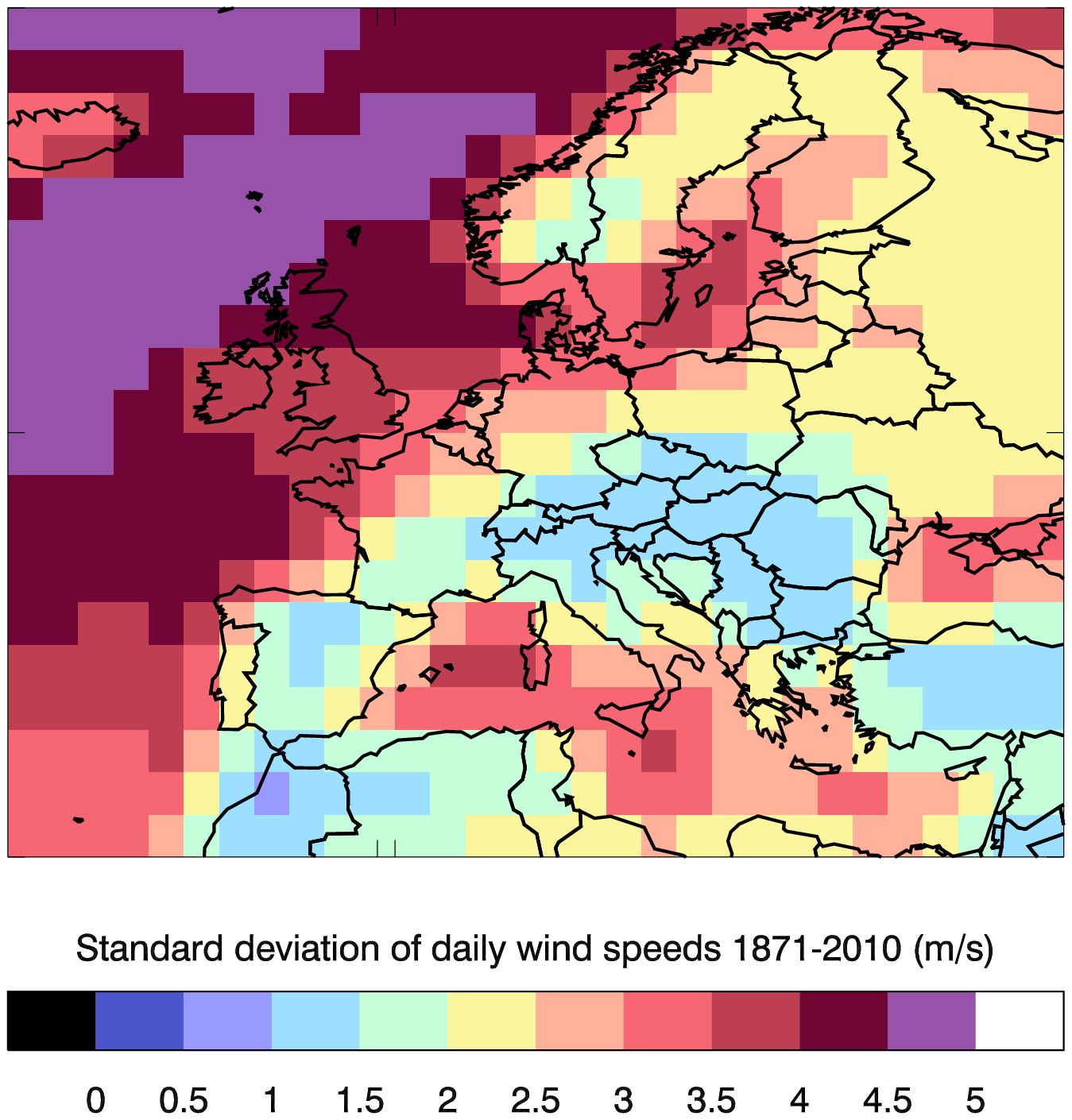}
  \end{center}
  \caption{Maps of the 140-yr mean and standard deviation of the
    daily wind speeds from 20CR.  The white box in the top plot marks
    the region we refer to as `eastern England'.}
  \label{f:windmaps_meanandvar}
\end{figure}

We show the resulting long-term maps in
Fig.~\ref{f:windmaps_meanandvar}.  The areas with the highest wind
speeds, and the greatest day-to-day variability, are over the sea
rather than the land.  In particular, the highest-wind regions are in
the north Atlantic, west of Ireland and north of Scotland.  Relatively
high winds also cover the rest of the British Isles and the coastal
areas of north-western Europe, as well as the central Mediterranean.

The extremely low winds over the areas of more complex orography (in
an arc from the Atlas Mountains, over Spain, the Alps, the
Carpathians, and through to Turkey) are likely to be due to the drag
scheme used to model the impact of unresolved orography. (Note that
the horizontal scale of the mountains is much smaller than the grid
size used here.)  While such features are often seen in maps of model
wind speeds \cite[e.g.][]{HowardClark2007}, it is important to be
particularly wary of the results in these areas, as it is not clear
how to relate them to the real physical situation `on the ground'.

\begin{figure} 
  \vspace*{2mm}
  \begin{center} 
    \includegraphics[width=0.8\columnwidth]{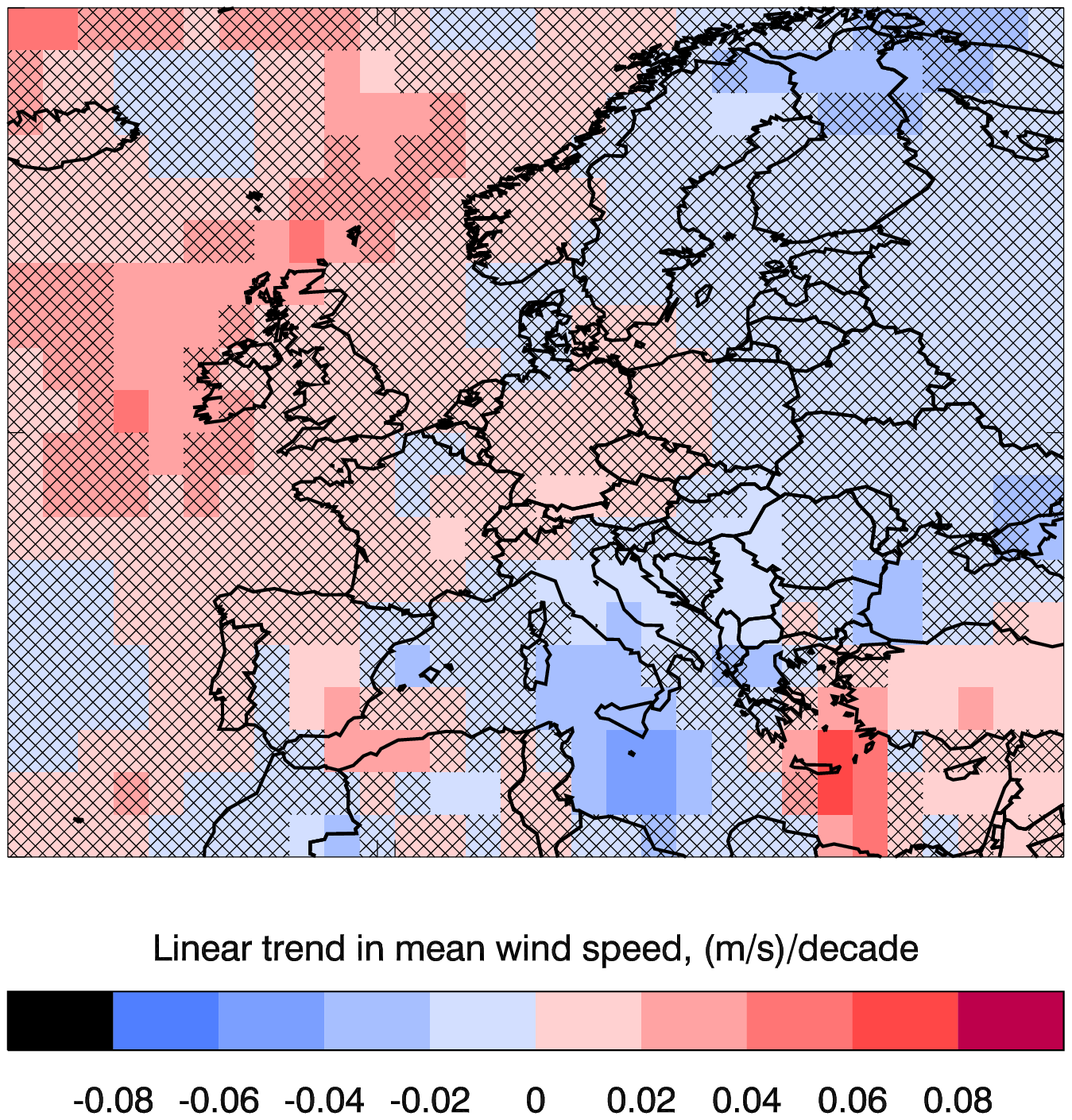}\\
    \includegraphics[width=0.8\columnwidth]{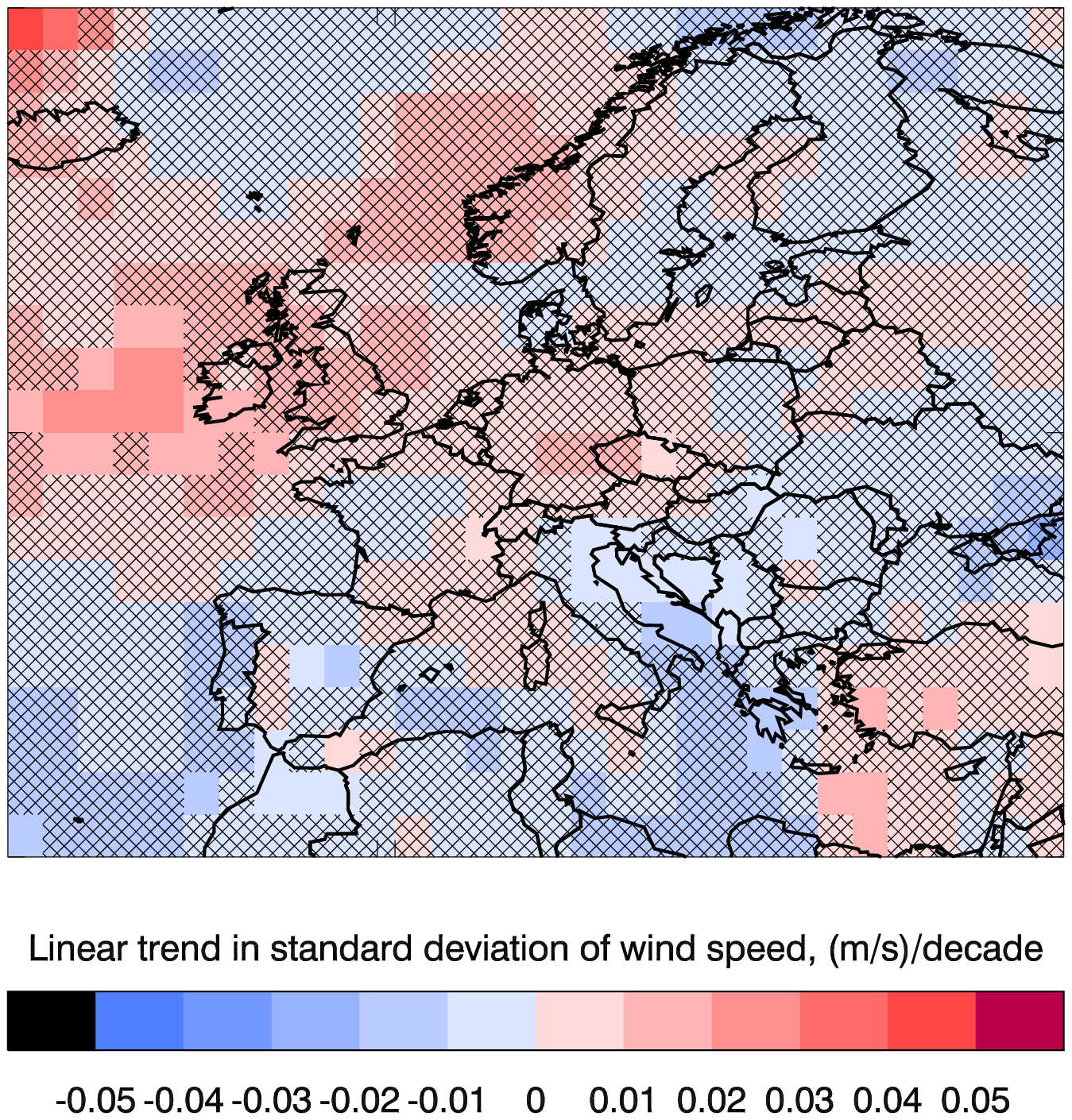}
  \end{center}
  \caption{Gradients of the linear regressions to the 140-yr time
    series of 5-yearly ensemble-mean average wind speeds (top) and
    standard deviations of daily wind speeds (bottom).  Shaded areas
    show where the trend is not statistically significant at the
    $0.1\%$ level.}
  \label{f:trendmaps_meanstdev}
\end{figure}

We have tested for the presence of any long-term trends, by performing
a simple linear regression on the ensemble-mean time series of 5-yr
means and standard deviations of the wind speeds in each grid cell.
The significance of any trend (i.e. whether the gradient of the linear
fit was significantly different from zero) was assessed using a
$t$ test at the $0.1\%$ level.  Figure~\ref{f:trendmaps_meanstdev} shows
the long-term trends, shaded out when they are not statistically
significant.

While most areas over Europe show no significant trend, there are some
areas of interest: the Atlantic off the north and west coasts of
Ireland and Scotland shows a significant positive trend (with the mean
wind speed increasing at a rate of $\sim 0.03\trendunit$), as does the
eastern Mediterranean around Crete and Turkey ($\sim
0.07\trendunit$). There is a significant negative trend in the central
Mediterranean around Sicily and Malta ($\sim -0.05\trendunit$).  Note,
however, that even where the trends are significant, they are still
extremely small: a change of $\sim \pm 1\windunit$ per two centuries.
The standard deviation trend map shows similar spatial patterns, but
even fewer and smaller areas where those trends are significant.

\begin{figure} 
  \vspace*{2mm}
  \begin{center} 
    \includegraphics[width=0.88\columnwidth]{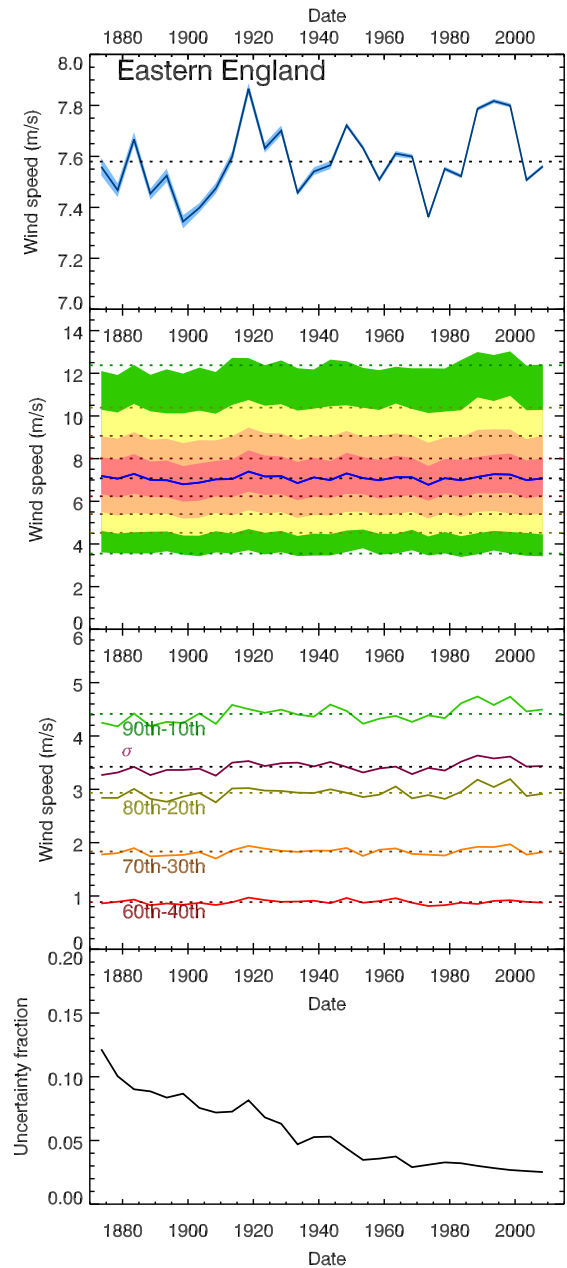}
  \end{center}
  \caption{Time series of the wind speed distribution averaged over
    the eastern England region.  Top: Ensemble mean of the 5-yr mean
    wind speed, with lighter shading indicating the 10th and 90th
    percentiles of the ensemble member distribution (most visible at
    early times).  Second from top: the ensemble mean of the 5-yr
    median wind speed (blue line), as well as the other deciles
    (coloured regions, from 10th percentile at the bottom to the 90th
    percentile at the top). The long-term means of the deciles are
    shown by dotted lines.  Widths of the distribution are shown in
    the panel second from bottom: Half the difference between
    ensemble-mean deciles (as labelled), and the ensemble mean of the
    standard deviation of ensemble members' wind speeds in 5-yr bins
    (labelled $\sigma$). Bottom: 5-yr means of the day-to-day
    ensemble spread (i.e. standard deviation of the ensemble members each
    day), as a fraction of the ensemble-mean 5-yr mean wind
    speed.  }
  \label{f:ts_eengl}
\end{figure}

We show an example time series in detail in Fig.~\ref{f:ts_eengl}, to
illustrate some general features of the wind time series over Europe,
and to show the benefits of being able to use such a long data set.
It uses daily wind speeds averaged over a region covering eastern
England (see box in Fig.~\ref{f:windmaps_meanandvar}).  This region
has been selected partly because it is an area of interest for future
wind farm development\footnote{See e.g. maps on the UK Government's
  RESTATS web site \url{http://restats.decc.gov.uk/app/pub/map/map/}
  and the UK Crown Estate's web site
  \url{http://www.thecrownestate.co.uk/energy-infrastructure/downloads/maps-and-gis-data/}},
but also because it avoids some of the more exceptional features
already discussed, such as the artificially low winds seen over
complex terrain and the weak trends seen further to the north-west.
We reserve a more detailed discussion of such features for subsequent
papers.

The 5-yearly mean wind speeds (top panel of Fig.~\ref{f:ts_eengl};
note the magnified scale) clearly show the peak in the late-1980s up
to around 2000, and the subsequent return to the long-term mean.  We
can also see that this excursion from the mean, while strong, is not
exceptional in the long-term history of this region (see e.g. the peak
in the 1915--1920 period, and the prolonged reduction over
1885--1910).  There is also no clear trend.

By looking at the deciles of the distribution (second panel in
Fig.~\ref{f:ts_eengl}), we can see how different features appear at
different levels.  For example, there is a clear trough in the early
1970s in the central region of the distribution; it doesn't appear in
the 10th and 90th percentiles, and only forms part of a much broader
reduction in the 80th percentile.  The length and form of the
reduction around the 1890--1910 period is also different in different
deciles, with the upper deciles showing a longer reduction in wind
speed. These features can be tracked through into the distribution
widths (third panel of Fig.~\ref{f:ts_eengl}): for example, the dip in
the widths over 1905--1910 is due to the lower deciles rising while
the upper deciles are still low.  The actual meteorological situations
in these periods would need to be investigated in more detail to
understand the underlying physical causes, but these features show the
importance of analysing the whole distribution beyond just the mean.

The day-to-day ensemble spread as a fraction of the mean wind speed is
shown in the bottom panel of Fig.~\ref{f:ts_eengl}.  This will always
be much greater than the uncertainty in the 5-yr mean shown in the
top panel; nonetheless, it illustrates a general decrease in the
uncertainty in the data, with an average daily spread between the
ensemble members of about 2.5\% of the mean wind speed by the 2000s.
This reduction in uncertainty is directly related to the dramatic
increase in the amount of data assimilated over the course of the 20CR
period, which constrains the ensemble members reducing their spread
(see discussions in e.g. \citealt{Compo2011, 2012JGRD..117.5123F,
  Krueger2012, Wang2012}).  It is important to note that this panel
shows the \emph{uncertainty} in the reanalysis (through the daily
ensemble spread), whereas the $\sigma$ in the panel above refers to
the \emph{variability} of wind speeds over time in all ensemble
members.


\subsection{The Weibull distribution}

\begin{figure*}[t]
  \vspace*{2mm}
  \begin{center} 
    \includegraphics[width=0.4\textwidth]{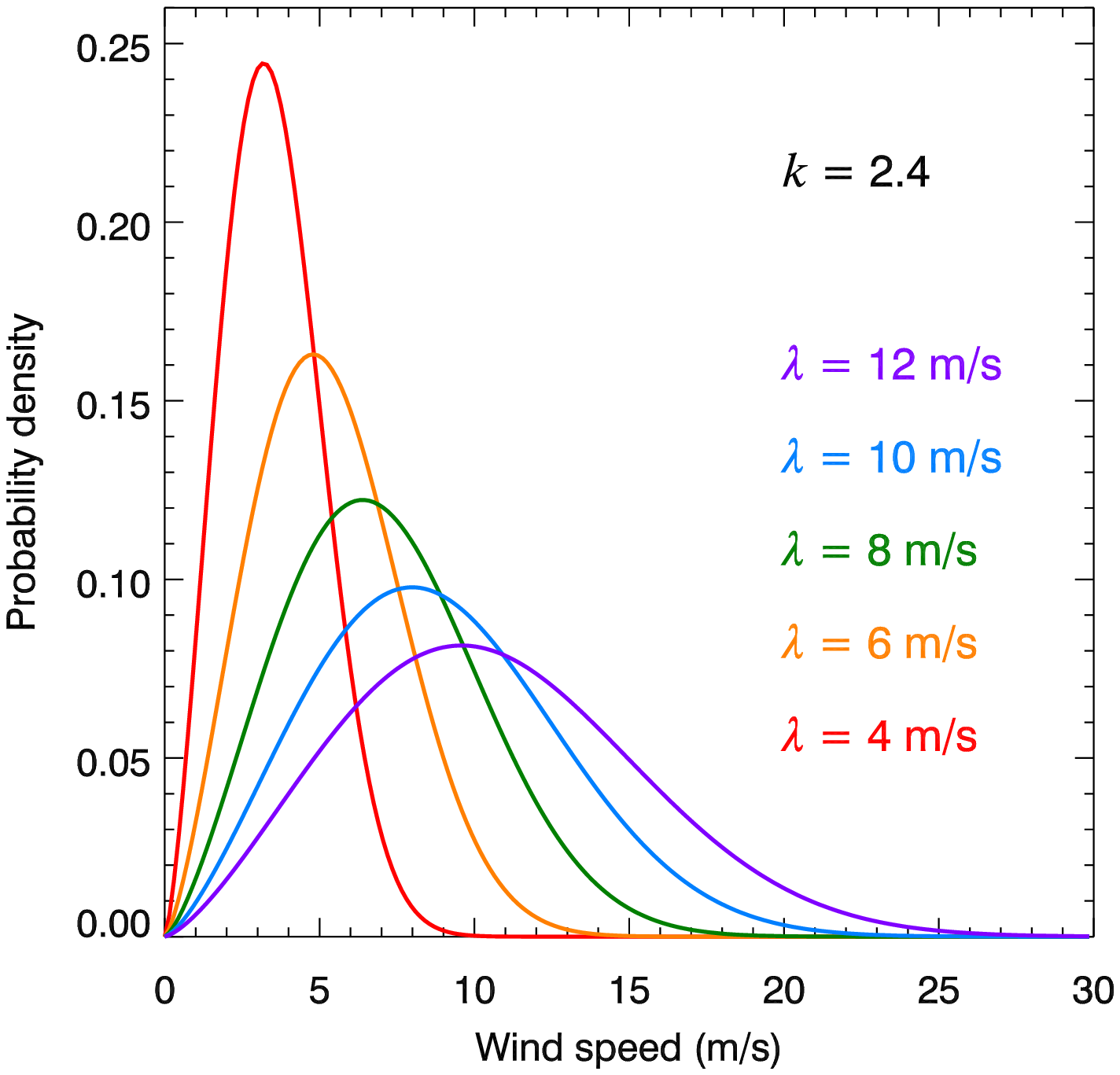}
    \includegraphics[width=0.4\textwidth]{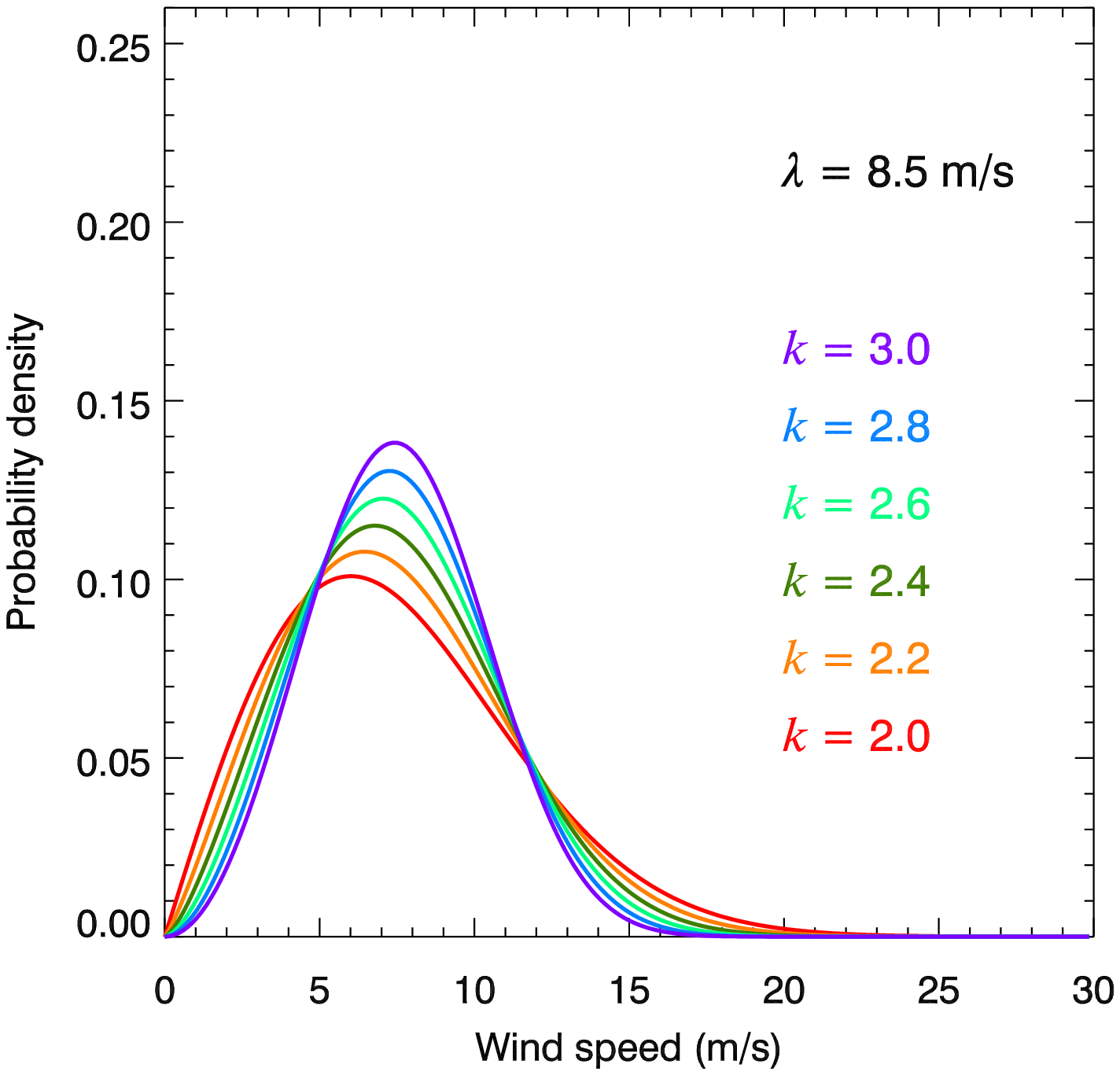}
  \end{center}
  \caption{Demonstrations of Weibull distribution PDFs when varying
    the scale parameter ($\lambda$, left) and the shape parameter
    ($k$, right).  Values chosen for display reflect those seen in our
    results.}
  \label{f:weibdemos}
\end{figure*}

Distributions of wind speeds at given locations are commonly modelled
as being drawn from a Weibull distribution \citep[e.g.][and references
therein]{Conradsen1984Review, Carta2009Review}, which has the
probability density function (PDF)
\begin{equation}
  \label{e:weibull}
  P(U) = \frac{k}{\lambda} \left( \frac{U}{\lambda} \right)^{k-1}
  \exp\left[ -\left( \frac{U}{\lambda} \right)^k \right],
\end{equation}
where the random variable $U$ is the wind speed, $k$ is the
(dimensionless) shape parameter, and $\lambda$ is the scale parameter
(in the same units as $U$).  The form of the distribution for
different values of $k$ and $\lambda$ is shown in
Fig.~\ref{f:weibdemos}.  The shape parameter $k$ determines the
overall form of the distribution, including the skewness, and the
width for a given $\lambda$; for wind speeds, $k$ tends to lie between
2 and 3.  For a given shape, the scale parameter $\lambda$ determines
the distribution's peak-location and width; higher values of $\lambda$
result in a broader distribution with a peak at a higher wind speed.
We can use the Weibull parameters as an alternative way of
characterising the wind speed distribution.  Examining changes in the
Weibull parameters allows us to track the behaviour of the
distribution as a whole, rather than e.g. just the mean value.

\begin{figure}[t]
  \vspace*{2mm}
  \begin{center} 
    \includegraphics[width=0.95\columnwidth]{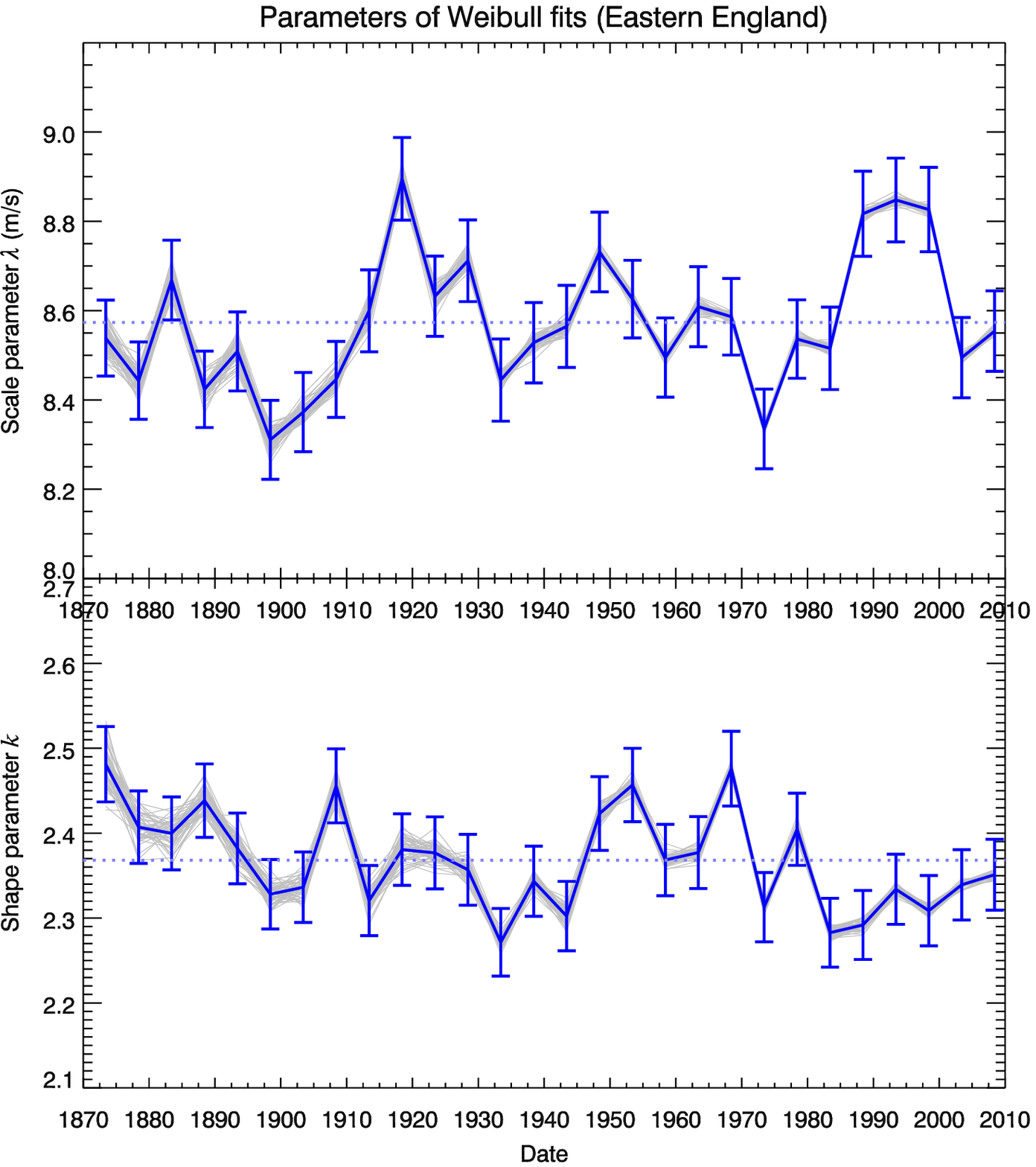}
  \end{center}
  \caption{Time series (in 5-yr bins) of the Weibull scale and shape
    parameters for our sample region covering eastern England.  The
    individual ensemble members are shown as grey lines (note that
    since ensemble members are discontinuous in these 5-yr steps,
    the lines do not reflect actual time series, but are plotted
    merely to guide the eye).  The blue line shows the ensemble-means
    of the Weibull parameters in each step, with the error bars given
    by the ensemble-means of the estimated standard errors from the
    fits.  The long-term means of the ensemble-mean time series are
    plotted as dotted lines.}
  \label{f:weibfitts_eengl}
\end{figure}

In a given grid cell or region, we find maximum-likelihood estimates
of the Weibull scale and shape for the distribution of daily wind
speeds of each ensemble member in each 5-yr period.  This yields
ensemble estimates of the time series of Weibull parameters; an
example from the same eastern England region as before is shown in
Fig.~\ref{f:weibfitts_eengl}.  As expected, the behaviour of the scale
parameter over time is similar to the mean and $\sigma$ time series
(Fig.~\ref{f:ts_eengl}), whereas this is not seen in the shape
parameter.  As the shape parameter governs the skewness of the
distribution, the variation in $k$ could be related to the presence of
extreme high-wind events, as these could cause the form of the
distribution to be fit better with a longer tail.  

\begin{figure}[t] 
  \vspace*{2mm}
  \begin{center} 
    \includegraphics[width=0.8\columnwidth]{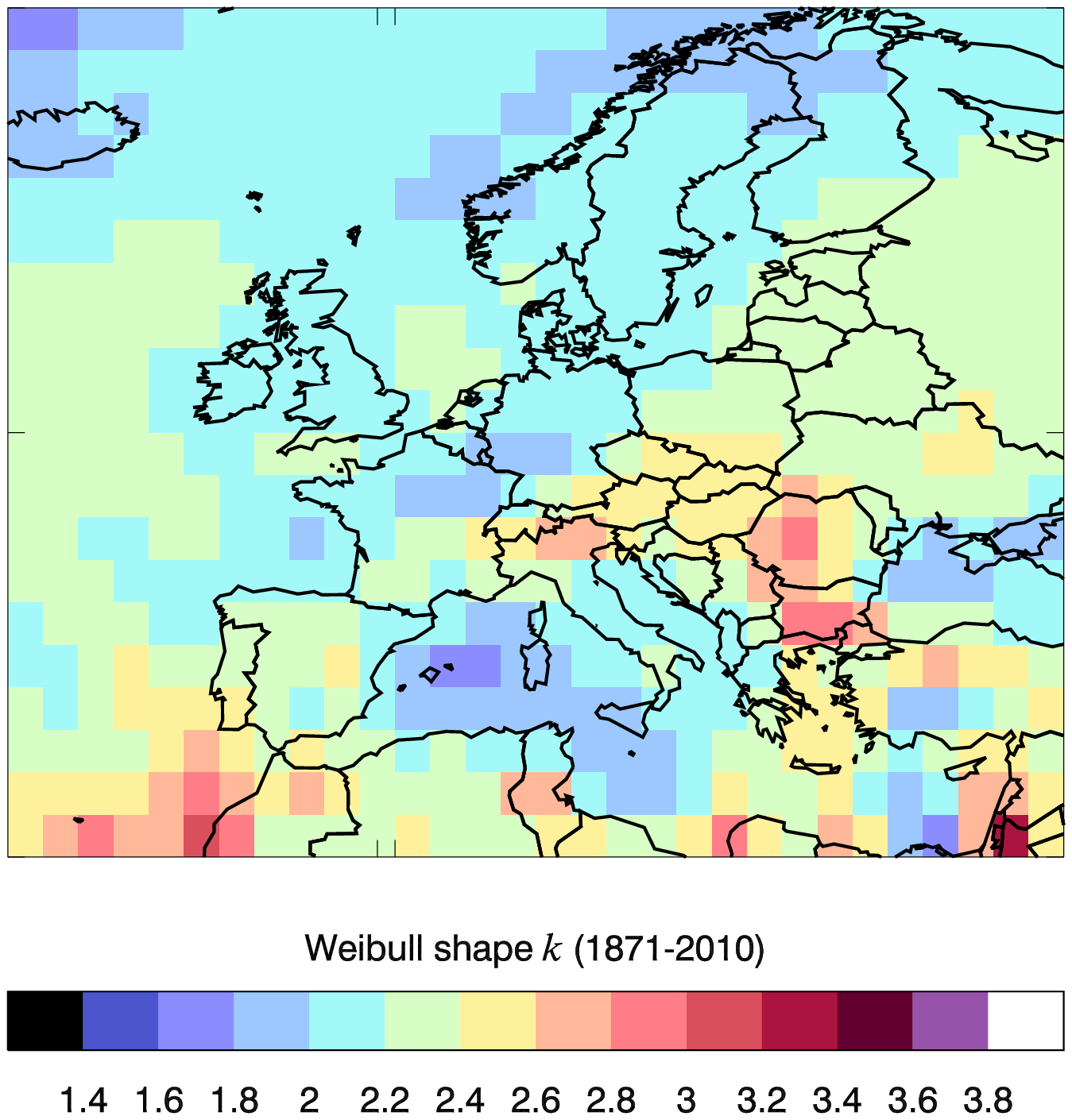}\\
    \includegraphics[width=0.8\columnwidth]{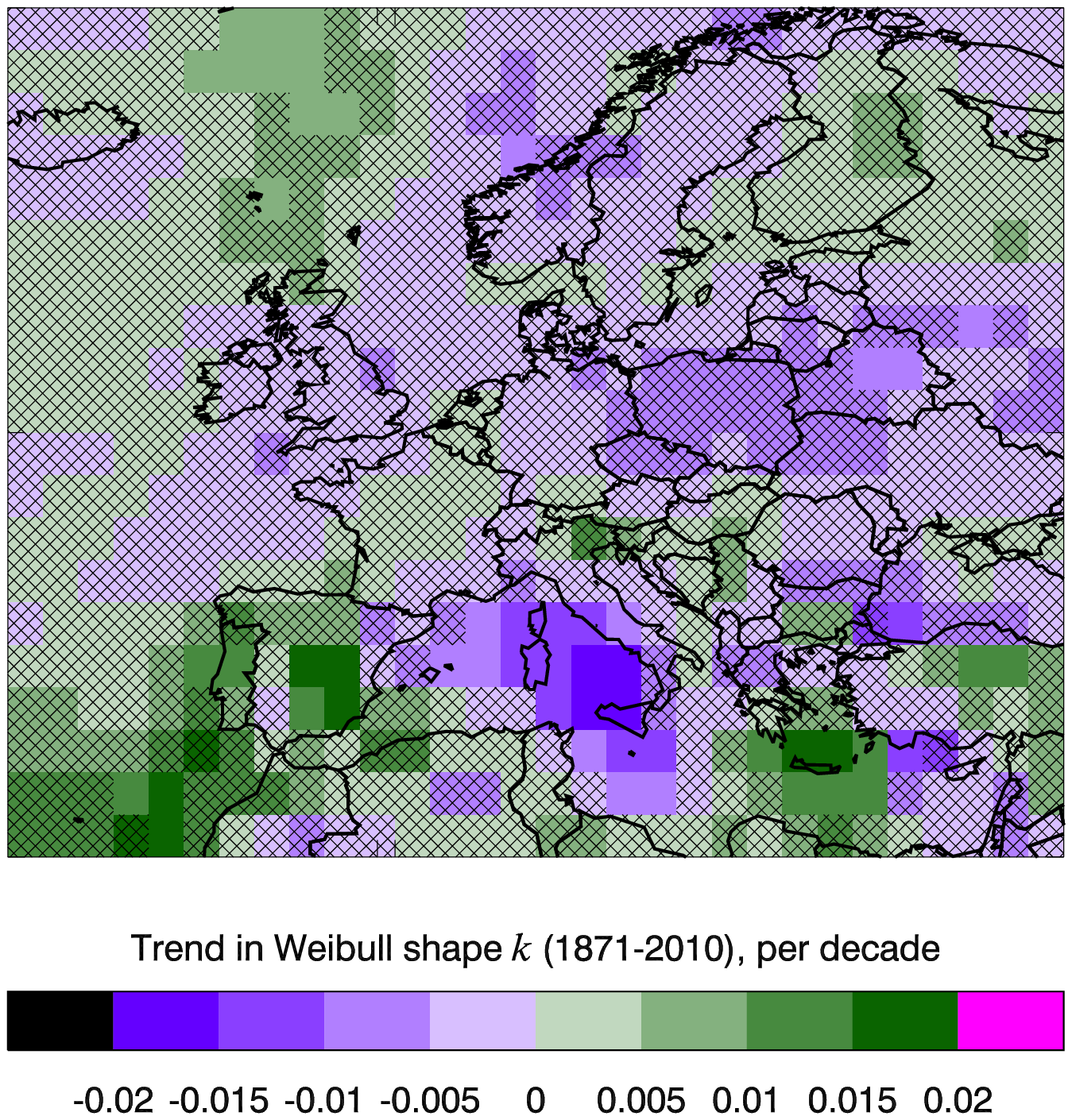}
  \end{center}
  \caption{Maps of the long-term mean of the ensemble-mean time series
    of Weibull shape parameters (top), and the trend from linear
    regresson on those time series (bottom). As before, areas where the
    trend is not significantly different from zero are shaded.}
  \label{f:weibfit_ltshapemaps}
\end{figure}

\begin{figure}[t] 
  \vspace*{2mm}
  \begin{center} 
    \includegraphics[width=0.8\columnwidth]{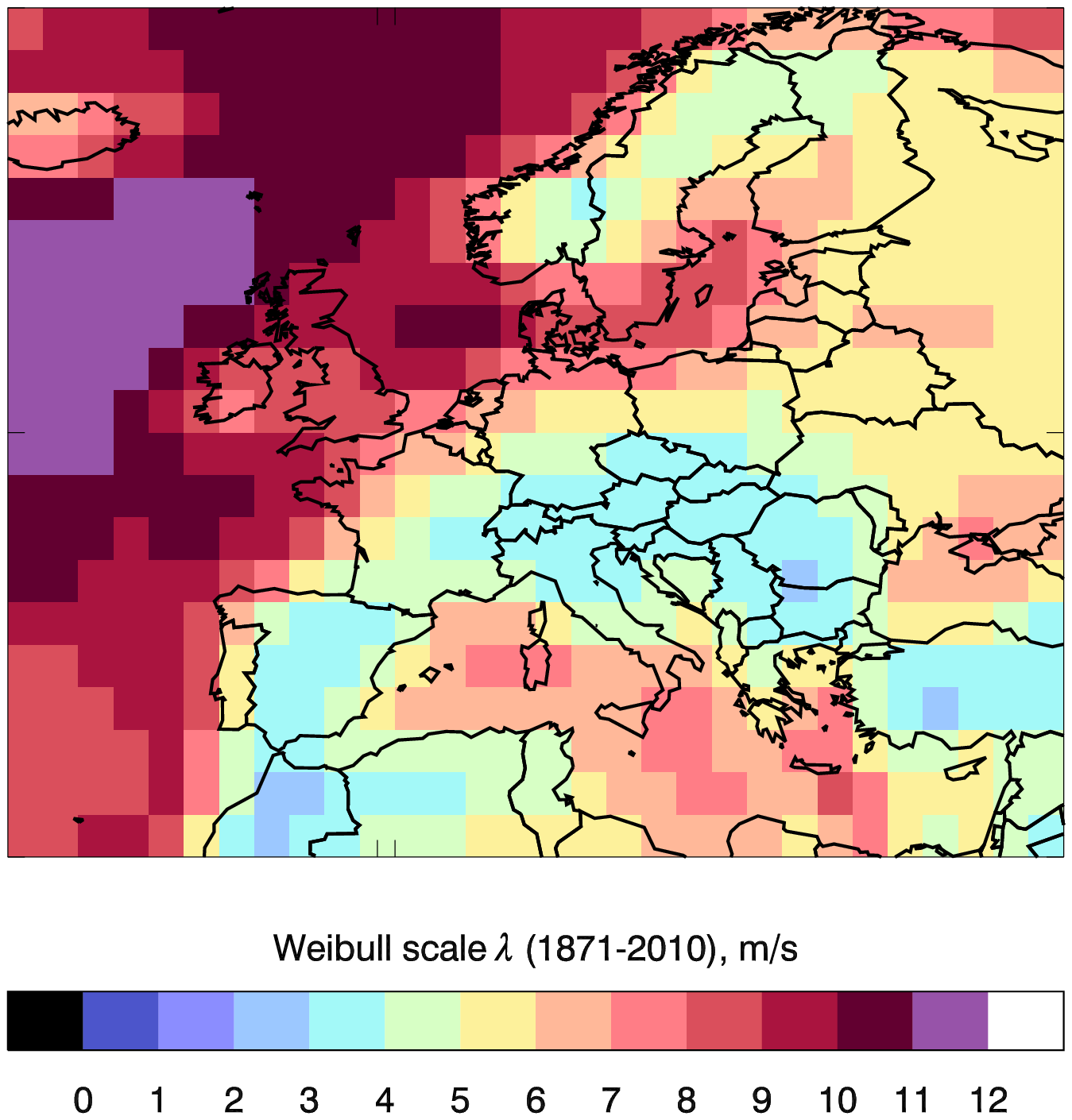}\\
    \includegraphics[width=0.8\columnwidth]{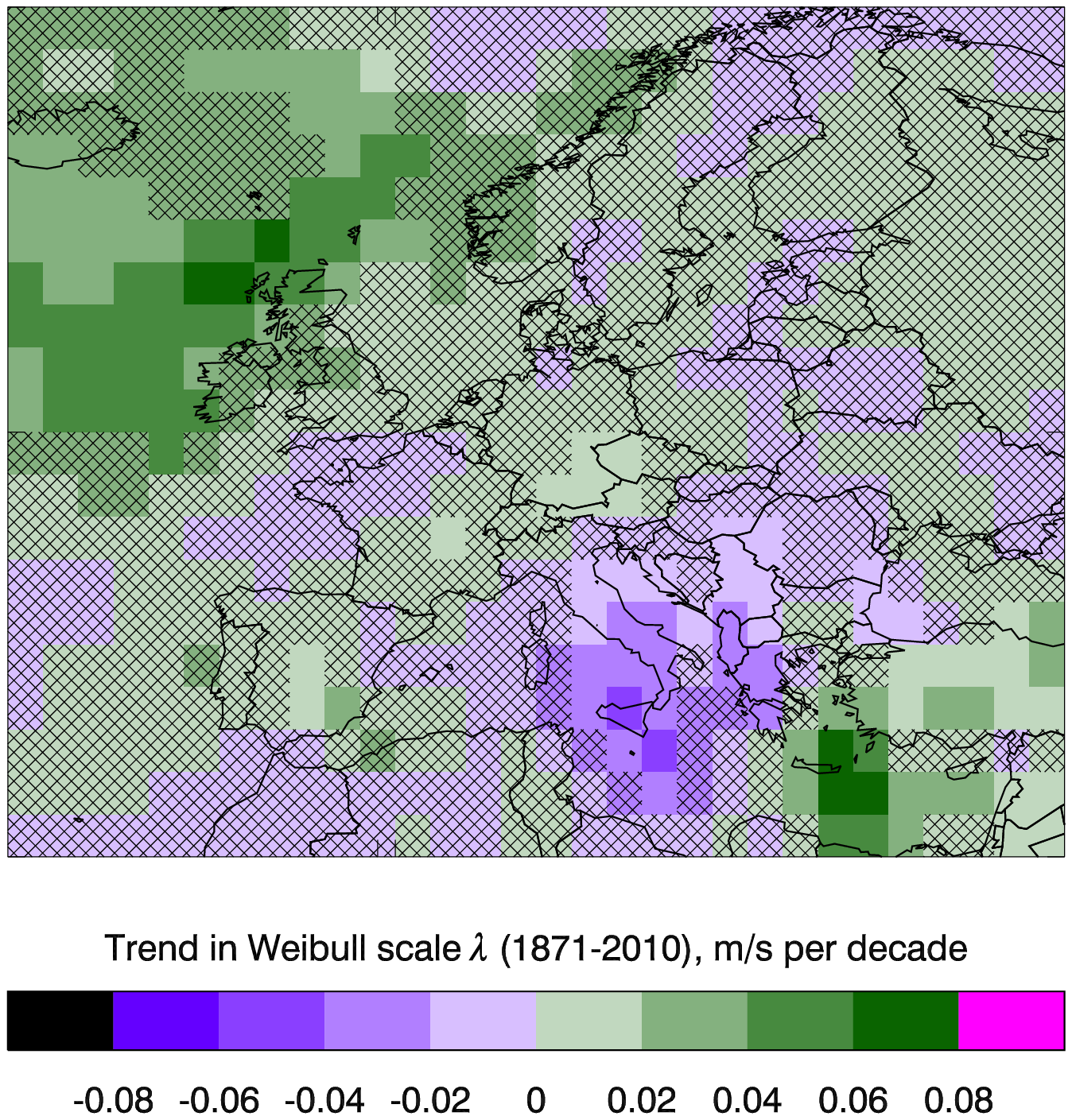}
  \end{center}
  \caption{As Fig.~\ref{f:weibfit_ltshapemaps}, but for the Weibull
    scale parameters.}
  \label{f:weibfit_ltscalemaps}
\end{figure}

Once again, we can extract linear trends from these time series, test
their significance, and map the results along with their long-term
values.  The long-term mean and trend of the shape parameter are shown
in Fig.~\ref{f:weibfit_ltshapemaps}.  The map for the average $k$
shows very little spatial variability, staying in most areas between 2
and 3.  Very few areas show any significant trend, although there is
again an area of reduction over the central Mediterranean
(i.e. tending to a more skewed distribution) and growth around Crete
(towards a more symmetric distribution).  The long-term mean and trend
of the scale parameter is shown in Fig.~\ref{f:weibfit_ltscalemaps}.
The mean-$\lambda$ map is very similar to both the mean and standard
deviation of the wind speeds (cf. Fig.~\ref{f:windmaps_meanandvar}),
and the trend map is again familiar; the area of significant growth in
the north-east Atlantic is in this case much larger.  

The distribution width is related to \emph{both} $\lambda$ and $k$,
but in opposite senses (e.g. narrower distributions can be obtained by
reducing $\lambda$ or increasing $k$, with corresponding impacts on
the peak locations and skewness).  Thus, it is not surprising that
(for example) the negative trends in both $k$ and $\lambda$ over Italy
and the Tyrrhenian Sea do not correspond with a significant trend in
the wind speed standard deviation
(cf. Fig.~\ref{f:trendmaps_meanstdev}).

It is important to note again that, regardless of statistical
significance, the trends present in both $k$ and $\lambda$ are still
exceedingly small.


\subsection{Year-ranking by wind speed}

\begin{figure*} 
  \vspace*{2mm}
  \begin{center} 
    \includegraphics[width=\textwidth]{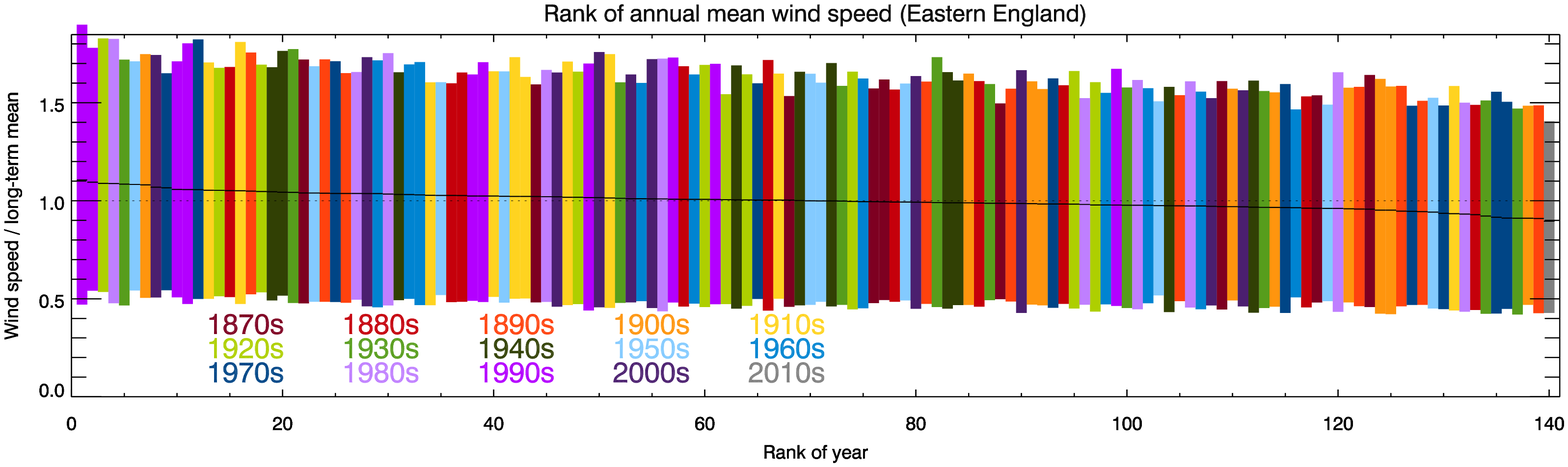}
  \end{center}
  \caption{Annual mean wind speeds (black line) in the eastern England
    region, divided by the 140-yr mean, and plotted in rank order.
    The vertical bars (one for each year, coloured by decade) show the
    10th and 90th percentile of the daily winds each year.}
  \label{f:ranks_spreads_eengl}
\end{figure*}

An additional way of relating the wind speeds in different years to
the long-term context is to plot the annual mean wind speeds in rank
order. This is shown qualitatively in
Fig.~\ref{f:ranks_spreads_eengl}, for the eastern England region.  We
can immediately see that the day-to-day variability in wind speed is
much larger than the year-to-year variability (in annual-mean wind
speed), and the mixture of colours shows the lack of a clear long-term
trend.

\begin{figure} 
  \vspace*{2mm}
  \begin{center} 
    \includegraphics[width=\columnwidth]{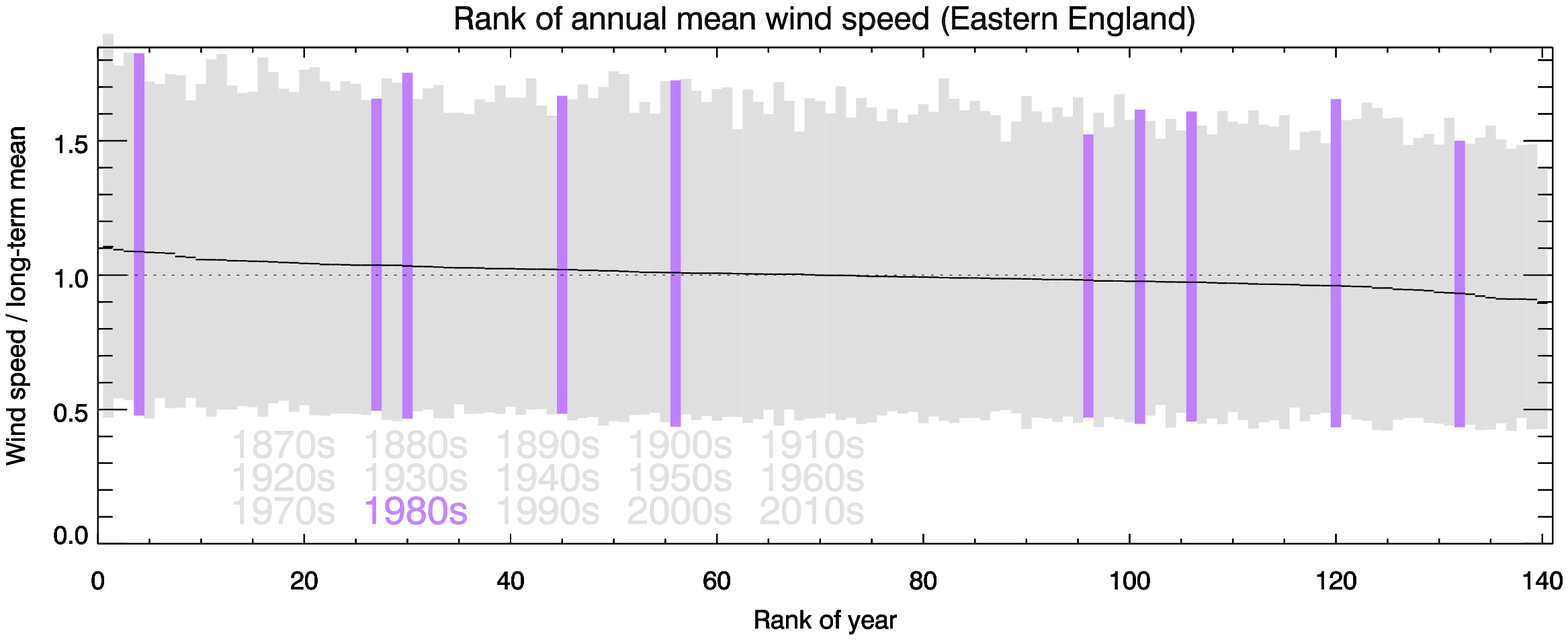}
    \includegraphics[width=\columnwidth]{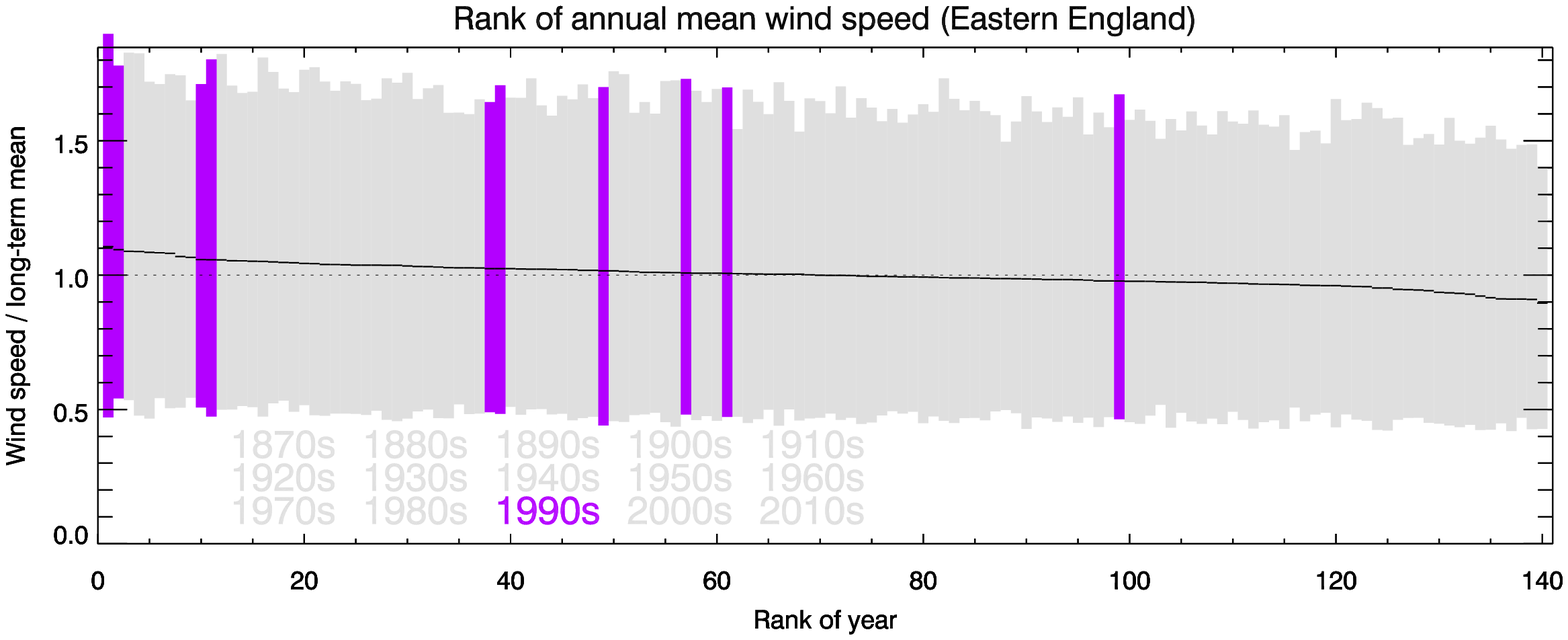}
    \includegraphics[width=\columnwidth]{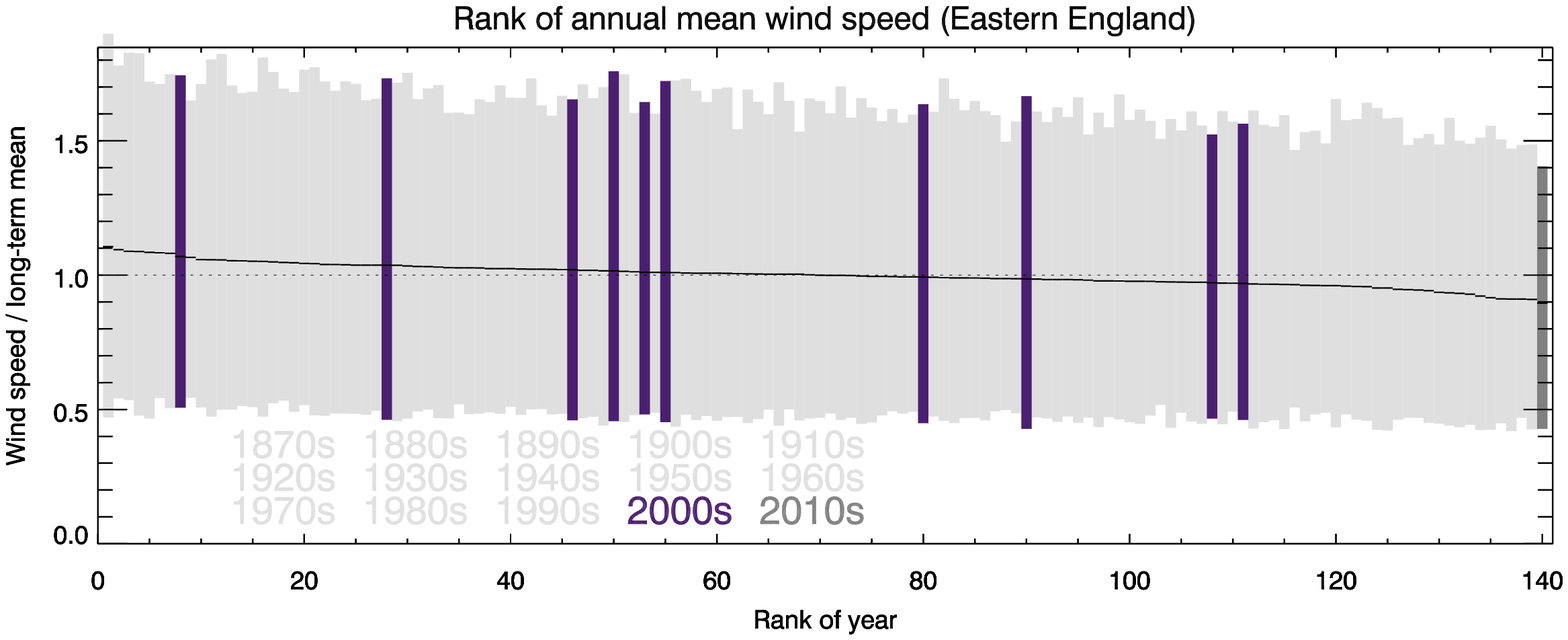}
  \end{center}
  \caption{As Fig.~\ref{f:ranks_spreads_eengl}, but highlighting the
    past three decades separately.  We have included 2010 as an
    additional year in the 2000s' plot. }
  \label{f:ranks_spreads_decades_eengl}
\end{figure}

We can get a more quantitative picture by highlighting individual
decades.  Fig.~\ref{f:ranks_spreads_decades_eengl} shows how the most
recent three decades are positioned with respect to the 140-yr time
series.  Here we can see the differences in the spreads of the
different decades: nine of the ten years of the 1990s (i.e. of the
years 1990--1999 inclusive) are in the top half of the distribution,
but only a couple are in the top 10.  The 1980s includes some very
high and very low wind years.  The 2000s are also well-spread (6 yr
in the top half, 4 yr in the bottom half; note 2008 is in the top
10), but it is 2010 that is the extreme low-wind year.  This is
because, since we are using calendar years, 2010 is hit by the low
winds (or lack of high winds) associated with the extreme cold
temperatures in January--February and November--December that year
\citep[e.g.][]{priorkendonukwinter0910, priorkendonlate2010}.

\conclusions[Summary and conclusions]

Using the Twentieth Century Reanalysis, we have analysed the
distribution of wind speeds over Europe, and how it has varied over
140 yr of daily data.  The regions with the highest long-term
average wind speeds, and highest day-to-day variabilities, are in the
north-east Atlantic, and the data exhibits a strong land--sea
contrast.  Most areas show no clear trends over time in the mean wind
speed or its variability; even in those areas where a trend is
statistically significant, the magnitude is so small that it is still
questionable whether it is physically meaningful or in any way
relevant for applications such as wind energy. Long-term trends are
difficult to distinguish from natural variability, and the variability
is much greater than the uncertainties in the data.

However, the decadal-scale variations in the long time series help us
understand the behaviour of recent decades in a long-term context.
For the region that we have illustrated as an example (covering
eastern England), we can see that the variations seen in wind speeds
in the past 20--30 yr are typical of the natural variability of
this region.  Indeed, the past thirty years do not cover the full
range of variability seen in the 140-yr time series, e.g. not
capturing the the lower-wind episodes of the 1970s and 1900s in
England.  Note that, while the results do not depend qualitatively on
the precise area in question, there is significant country-scale
variability across Europe, both quantitatively and qualitatively.

Overall, we have shown that long-term data sets are essential in
understanding wind speed distributions across Europe.  In subsequent
papers, we will describe our analysis and results in more detail,
including a comparison with other data sets.








\begin{acknowledgements}
  PB acknowledges the support of an EMS Young Scientist Travel Award.
  Support for the Twentieth Century Reanalysis Project dataset is
  provided by the U.S. Department of Energy, Office of Science
  Innovative and Novel Computational Impact on Theory and Experiment
  (\href{http://science.energy.gov/ascr/facilities/incite/}{DOE INCITE})
  program, and Office of Biological and Environmental Research
  (\href{http://science.energy.gov/ber/}{BER}), and by the
  National Oceanic and Atmospheric Administration
  \href{http://www.climate.noaa.gov/}{Climate Program Office}. 

The works published in this journal are distributed under the
Creative Commons Attribution 3.0 License. This license does not
affect the Crown copyright work, which is re-usable under the Open
Government Licence (OGL). The Creative Commons Attribution
3.0 License and the OGL are interoperable and do not conflict with,
reduce or limit each other.

\copyright Crown copyright 2013

Edited by: S.-E. Gryning\\
Reviewed by: O. S. Rathmann and one anonymous referee
\end{acknowledgements}


\bibliographystyle{copernicus}
\bibliography{philipbett_ems2012_article}

\end{document}